\documentclass[aip,graphicx]{revtex4-1}

\usepackage[T2A]{fontenc}
\usepackage[english]{babel}
\usepackage{amsmath} 
\usepackage[%
	breaklinks=true,
	colorlinks=true,
	linkcolor=blue,
	urlcolor=blue,
	citecolor=blue
]{hyperref}
\usepackage{graphicx}

\draft 

\begin{document}


\title{Simulation of the thermocapillary assembly of a colloidal cluster during the evaporation of a liquid film in an unevenly heated cell} 



\author{Kristina N. Kondrashova}
\affiliation{Astrakhan Tatishchev State University, Russia}

\author{Konstantin S. Kolegov}
\email[]{konstantin.kolegov[at]asu-edu.ru}
\affiliation{Astrakhan Tatishchev State University, Russia}

\author{Irina V. Vodolazskaya}
\affiliation{Astrakhan Tatishchev State University, Russia}



\date{\today}

\begin{abstract}
The control of the thermocapillary assembly of colloidal particle clusters is important for a variety of applications, including the creation of photonic crystals for microelectronics and optoelectronics, membrane formation for biotechnology, and surface cleaning for laboratory-on-chip devices. The aim of the current work is to understand the main mechanisms influencing the assembly of such clusters, which are the object of our research. Here we consider a two-dimensional mathematical model describing the transfer of particles by a thermocapillary flow in an unevenly heated cell during the evaporation of a liquid. This gave us the opportunity to study one of the main processes that triggers the formation of a particle cluster. Whether the particle will move with the flow or stop at the heater, becoming the basis for the  cluster, is determined by the ratio between gravity and the drag force.  The results of numerical calculations show that, for small particle concentrations, their fraction entering the cluster decreases as the volumetric heat flux density $Q$ increases.  The reason for this is an increase in the thermocapillary flow with an increase in the volumetric heat flux $Q$. In our conclusion, it reduces the probability of particles entering the cluster.
\end{abstract}


\maketitle 

\section{Introduction}
\label{sec:Introduction}
Evaporation of droplets and films is often accompanied by the formation of various types of colloidal structures~\cite{Kolegov2020}. One way to control this process is through uneven spatial heating of the substrate. The temperature gradient of the solid surface affects hydrodynamics in the liquid layer. Thus, it is possible to influence the transfer of colloidal particles.

In experiment ~\cite{Malla2019} the effect of particle size and the temperature gradient of the substrate on the effect of coffee rings was studied. In room conditions, an annular deposit formed in place of the dried droplet due to particle motion under the action of capillary flow towards the contact line. The particle size affected the contact line mode (``pinning'' or ``stick--slip motion''). The temperature gradient of the substrate led to the appearance of the Marangoni flow, which also affected the transfer of particles. This led to an asymmetry in the shape of the deposit or its different geometry~\cite{Malla2019}. An increase in the substrate temperature relative to the ambient temperature led to an increase in the width of the annular deposit (with the presence of a central depleted zone) up to the formation of a continuous spot~\cite{Lama2020}. In another experiment with a droplet, with increasing substrate temperature, a transition from a precipitate in the form of a ``coffee ring'' to a ``coffee eye'' was observed. Last one geometry included a narrow annular deposit and a massive central spot inside it~\cite{Li2015}. The simulation results showed that when the substrate temperature increases, the same change in the geometry of the deposit occurs~\cite{Zhang2021}. Numerical simulations and experiments were conducted for the case of partial heating of a substrate, where a drop of water containing polystyrene particles was clamped in a Hele-Shaw cell between side plates positioned perpendicular to the horizontal surface of the substrate~\cite{Thokchom2015}. This cell shape made it possible to simplify the problem and move on to considering it in 2D. For different heating modes, the flow structure was visualized by the PIV method and calculated using the equations of hydrodynamics. In addition, experiments were conducted on 3D droplets to obtain images of the deposit geometry. Based on these images, particle concentration profiles (the proportion of particles in a local subdomain relative to the total number of particles) were plotted relative to the radial coordinate for different heating modes~\cite{Thokchom2015}. The effect of substrate temperature and concentration of aluminum oxide nanoparticles on deposit geometry at site of dried droplet was studied experimentally~\cite{LiuB2020}. At a certain range of values for the key parameters, a formation of deposit in the shape of a double ring was observed. This deposit includes a ``coffee ring'' and an inner ring~\cite{LiuB2020}. A binary mixture of particles of different sizes was used in an experiment with a drop drying on a heated substrate~\cite{Parsa2017}. Depending on the temperature of the substrate, five different deposition patterns could be observed: a relatively uniform pattern enclosed by a disk-shaped ring, a nearly non-uniform pattern inside a thick outer ring, a ``dual-ring'' pattern, a ``rose-like'' pattern, and a set of concentric rings corresponding to ``stick--slip'' pattern. The annular deposit in the case of a binary mixture consisted of two sub-rings separated by a small gap. The outer ring consisted of small particles, while the inner ring usually consisted of large particles~\cite{Parsa2017}. The authors~\cite{Patil2018} conducted similar experiments and additionally studied the effect of a parameter related to the ratio of the diameter of small particles to that of large ones, $d_s / d_l$.  The resulting phase diagram showed at what values of the parameters $T_s$ and $d_s$/$d_l$ self-sorting of particles by size within the ring occurs~\cite{Patil2018}. For uneven heating of the open cell with liquid, a copper rod was connected to the Peltier element and mounted in the central part of the bottom of the cell~\cite{AlMuzaiqer2021}. The resulting temperature gradient on the free surface of the liquid led to the appearance of a thermocapillary fluid flow. The transfer of particles by the flow influenced the creation of a cluster of particles in the central region of the cell. The numerically calculated flow velocity in order of magnitude was in good agreement with the measurement results obtained by the PIV method~\cite{AlMuzaiqer2021}. The proposed 1D model allows to consider only the horizontal flow velocity averaged over the height of the liquid film. But it doesn't allow one to determine the 2D structure of the flow~\cite{AlMuzaiqer2021}. The simulation showed that, over time, the concentration of particles increases in the area of the heater. This technology can be used to control microparticles, including solids, polymers, cells, micelles, proteins, and microemulsions that are suspended in a liquid medium (see Refs. in paper~\cite{AlMuzaiqer2021126550}). The results of numerical calculations of the flow of pure liquid (without particles) in an evaporating droplet predicted the possibility of controlling the flow structure by changing the point of local heating of the substrate~\cite{Lee2022}. Another numerical study~\cite{Jain2024} showed that at temperature $T_s =$ 25$^\circ$C capillary flow prevails in the evaporating droplet. The thermal Marangoni flow dominates at temperature $T_s =$ 50$^\circ$C. In addition, the effect of the ratio of particle density to liquid density, $\rho_p / \rho_l$, on particle motion in a fluid flow was studied~\cite{Jain2024}. The effect of the nanoparticle concentration on the structure of deposits remaining after droplet drying on a heated substrate ($T_s =$ 80$^\circ$C) was studied both numerically and experimentally~\cite{Kumar2021}. At a relatively low concentration of the solution, an entire spot was formed. When the concentration of nanoparticles was relatively high, the coffee ring effect was observed~\cite{Kumar2021}. In Ref.~\cite{Goncharova2024}, the effect of two heaters mounted in a substrate on the structure of the flow of pure liquid in an evaporating film was studied using the proposed mathematical model. The corresponding flow structure was shown for different heating modes. The effect of particle sedimentation on the precipitation structure after drying of the pendant droplet on substrates of different wettability was investigated in Ref.~\cite{Kumar2020}. The obtained precipitation structures were compared with the structures for the case of a sessile drop. By adjusting the ambient temperature, it is possible to control the formation of an array of microneedles when the polymer solution film dries~\cite{Mansoor2011}. Such devices are useful for transdermal drug delivery system in medicine.

In this work, the research started in  Refs.~\cite{AlMuzaiqer2021126550,AlMuzaiqer2021} continues. We consider the case of a heater mounted into the bottom of an open cell. The purpose of this work is to study the influence of volumetric heat flux density on emerging clusters and to understand the mechanisms that lead to the appearance of a central cluster.

\section{Methods}
\label{sec:Methods}
\subsection{Physical problem statement}

\begin{figure}
	\centering
	\includegraphics[width=0.6\textwidth]{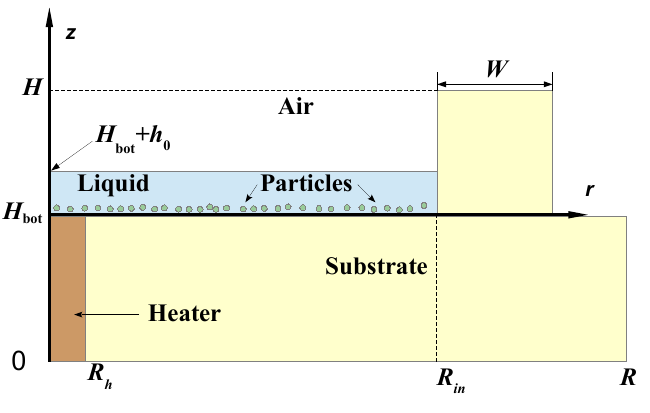}
	\caption{Sketch of the problem domain.}
	\label{fig:SketchOfCellWithHeater}
\end{figure}

Let us consider a liquid located in an open cylindrical cell, in the lower part of which a heater is mounted (Fig.~\ref{fig:SketchOfCellWithHeater}). The cell consists of a substrate of plexiglass, with a photopolymer ring glued on top of it. A copper rod is mounted in the center of the bottom of the cell, which is connected to the heated side of a Peltier element. As working materials, polystyrene microspheres and volatile isopropyl alcohol (isopropanol) were used. The problem has axial symmetry and is solved in a cylindrical coordinate system $(r,\,z)$. The geometrical parameters and properties of materials are summarized in Table~\ref{tab:GeometricParameters} and ~\ref{tab:PhysicalParameters}, their values are taken from Ref.~\cite{AlMuzaiqer2021}

\begin{table}
	\caption{Geometric parameters}
	\centering
	\begin{tabular}{|p{0.12\linewidth}|p{0.45\linewidth}|p{0.15\linewidth}|}
		\hline
		Symbol& Parameter& Value, mm\\
		\hline
		$h_0$&  Initial film height & 0.7\\
		$r_p$&  Particle radius &  0.025\\
		$H_{bot}$& Cell bottom thickness & 3 \\
		$R_h$& Heater radius & 0.6 \\
		$R_\mathrm{in}$&  Inner radius of the cell & 10\\
		$W$& Cell wall thickness & 4\\
		$H$& Cell wall height & 4\\
        $R$& Substrate radius & 19.5 \\
		\hline
	\end{tabular}
	\label{tab:GeometricParameters}
\end{table}

\begin{table}[h!]
	\caption{Physical parameters~\footnote{The indices $l$, $s$ and $p$ show the relationship of a parameter with the liquid, solid (the glass cell), and  particles, respectively. The parameter values are specified for the temperature 20$^{\circ}$C.}}
	\centering
	\begin{tabular}{|p{0.12\linewidth}|p{0.58\linewidth}|p{0.22\linewidth}|}
		\hline
		Symbol&  Parameter& Value\\
		\hline
		$L$& Isopropanol heat of vaporization & $75\times 10^4$ J$/$kg \\
		$M$& Isopropanol molar mass & 0.06  kg$/$mol\\
		$\rho_l$& Isopropanol density & 786 kg$/$m$^3$\\
		$\rho_s$& Plexiglass density & 2500  kg$/$m$^3$ \\
		$\rho_p$& Polystyrene particle density & 1050  kg$/$m$^3$\\
		$\rho_\mathrm{rod}$& Copper density & 8960 kg$/$m$^3$\\
		$P_{\nu}$& Partial pressure of isopropanol vapors in the air & $4.23\times 10^3$  Pa\\
		$T_0$& Room temperature & 298.15  K\\
		$T_\mathrm{sat}$&  Isopropanol vapors saturation temperature & 355.4 K\\
		$\mu$& Isopropanol dynamic viscosity& $2.43 \times 10^{-3}$  Pa s\\
		$c_l$& Isopropanol specific heat capacity & 2605 J$/$(kg K)\\
		$c_s$& Plexiglass specific heat capacity & 780 J$/$(kg K)\\
		$c_\mathrm{rod}$& Copper specific heat capacity & 385 J$/$(kg K)\\
		$\kappa_l$& Isopropanol thermal conductivity & 0.13 W$/$(m K)\\
		$\kappa_s$& Plexiglass thermal conductivity & 0.748 W$/$(m K)\\
		$\kappa_\mathrm{rod}$& Copper thermal conductivity & 400 W$/$(m K)\\
		$\alpha_\mathrm{air}$& Air heat transfer coefficient & 5.6 W$/$(m$^2$ K)\\
		$\alpha_l$& Isopropanol heat transfer coefficient &216 W$/$(m$^2$ K)\\
		$\alpha_s$& Plexiglass heat transfer coefficient & 2.5 W$/$(m$^2$ K)\\
		$\sigma_0$&  Surface tension & $22\times 10^{-3}$  N$/$m\\
		$\sigma^\prime$& Isopropanol surface tension temperature coefficient, $\sigma^\prime= \partial \sigma / \partial T$ & $-8\times 10^{-5}$  N$/$(m K)\\
		\hline
	\end{tabular}
	\label{tab:PhysicalParameters}
\end{table}
The dynamic viscosity of the liquid, $\mu$, and the density, $\rho_l$, are considered constant values, as we consider a small concentration of particles. We also consider the liquid to be incompressible.

\subsection{Mathematical model}

To calculate the flow velocity of a liquid, $\vec{v} = (v_r,\,v_z)$, let's use the continuity and Navier--Stokes equations,
\begin{equation}\label{ContinuityEquation}
\mathrm{div}\, \vec{v} = 0,
\end{equation}
\begin{equation}\label{eq:NavierStokes}
	\frac{\partial \vec{v}}{\partial t}+\left(\vec {v},\vec{\bigtriangledown} \right)\vec{v}=-\frac{1}{\rho_l}\mathrm{grad} \, p+\frac{\mu}{\rho_l}\Delta \vec{v},
\end{equation}
where $p$ is the pressure.

The propagation of thermal energy in a liquid medium is described using the heat transfer equation
\begin{equation}
	\rho_l c_l \left(\frac{\partial T}{\partial t}+\left(\vec {v},\vec{\bigtriangledown} \right)T \right) =\mathrm{div} \, \left(\kappa_l \, \mathrm{grad} \,T\right).
	\label{eq:HeatTransferLiquid}
\end{equation}\label{eq:HeatTransferInCell}
We will also write down the thermal conductivity equation for the glass cell,
\begin{equation}
	\rho_s c_s \frac{\partial T}{\partial t} =\mathrm{div} \, \left(\kappa_s \, \mathrm{grad} \,T\right),
\end{equation}
and for a copper rod,
\begin{equation}\label{eq:HeatTransferInRod}
	\rho_\mathrm{rod} c_\mathrm{rod} \frac{\partial T}{\partial t} =\mathrm{div} \, \left(\kappa_\mathrm{rod} \, \mathrm{grad} \,T\right) + Q,
\end{equation}
where $Q$ is the volumetric heat flux density (W/m$^3$). This model uses a single temperature notation $T$ for each subdomain, since Eqs.~\eqref{eq:HeatTransferLiquid}--\eqref{eq:HeatTransferInRod} can be rewritten as a single general equation, where the parameter values change depending on the subdomain for which it is being calculated. Furthermore, the convective term contained in Eq.~\eqref{eq:HeatTransferLiquid} and the source term $Q$ from Eq.~\eqref{eq:HeatTransferInRod} can automatically be set to zero in those subdomains where they are not required.

Suppose that the particles are small, their concentration is low and they do not affect the flow of the liquid. This makes it possible to independently calculate the velocity field of the liquid, $\vec{v}$, and the velocities of the particles, $\vec{u}$. The motion of a particle with mass $m_p$ is determined by Newton's second law,
\begin{equation}\label{eq:SecondNewtonLaw}
	m_p \frac{d \vec{u}}{dt} = \vec{F}_d + \vec{F}_g + \vec{F}_A ,
\end{equation}
where the drag force is $\vec{F}_d = m_p  \left(   \vec{v} - \vec{u} \right)/ \tau_p$, particle response time is  $\tau_p = \rho_p d_p^2/(18 \mu)$, particle diameter is $d_p = 2 r_p$, the gravity force is $\vec{F}_g = m_p \vec{g}$ ($\vec{g}$ is the acceleration due to gravity), the buoyant force (Archimedes' principle) is $\vec{F}_A = - \rho_l V_p \vec{g}$, and particle volume is $V_p = 4 \pi r_p^3/3$. In Eq.~\eqref{eq:SecondNewtonLaw}, we neglect the added mass, since the Stokes number $\mathrm{St}=\tau_p / \tau_f \approx 3\cdot 10^{-4} \ll 1$. Here, the characteristic flow time is calculated by the formula $\tau_f = \rho_p h_0^2 / \mu \approx$ 0.2~s.

\subsection{Initial and boundary conditions}

At the initial moment, the particles are randomly distributed near the bottom of the cell due to sedimentation ($\rho_p/ \rho_l \approx 1.34$, see Fig.~\ref{fig:SketchOfCellWithHeater}). The flow velocity and the particle velocity are equal to zero. The temperature of the liquid and the cell temperature are equal to the ambient temperature $T_0$, the free surface of the liquid is flat:
$$ \vec{v}=0, \quad \vec{u}=0, \quad T = T_0, \quad h=h_0.$$
The bottom boundary of the substrate is thermally insulated, $\partial T/\partial n = 0, \quad \vec{n}$ is the normal to the boundary. At the ``cell--air'' boundary, heat exchange with the environment is taken into account using the Newton--Richman law,
$$
	\kappa_{s} \frac{\partial T}{\partial n}=\lambda_{s} \left(T_0 - T  \right), \qquad \lambda_{s,l} = \left( \frac{1}{\alpha_{s,l}} + \frac{1}{\alpha_\mathrm{air}}  \right)^{-1}.
$$
In this empirical regularity $\lambda_{s,l}$ is the heat transfer coefficient. On the free surface of the liquid, heat flow associated with loss of thermal energy due to evaporation is also considered,
$$
	\kappa_l \frac{\partial T}{\partial n}=\lambda_l \left(T_0 - T  \right) - L J,
$$
where $J$ is the vapor flux density. There is no heat flux through the axis of symmetry (OZ axis), $\partial T / \partial r = 0$. At the ``liquid--glass'', ``liquid--copper'', and ``glass--copper'' boundaries, we take into account the equality of temperatures and heat fluxes: $$T_{l,\mathrm{rod},s} = T_{\mathrm{rod},s,l},\, \kappa_{l,\mathrm{rod},s} \frac{\partial T_{l,\mathrm{rod},s} }{ \partial n} = \kappa_{\mathrm{rod},s,l} \frac{\partial T_{\mathrm{rod},s,l} }{\partial n}.$$ Here, the subscripts for $T$ are used solely for ease of perception and understanding of the physical meaning of the boundary conditions. It should be remembered that there is only one temperature function in the model.

The model takes into account the velocity of the liquid free boundary motion $\vec{v}_\mathrm{mesh}$ caused by the fluid flow and evaporation. The boundary velocity component normal to the surface is expressed using the arbitrary Lagrangia--Eulerian (ALE) method,
$$
	\vec{v}_\mathrm{mesh} \cdot \vec{n} = \left( \vec{v} - \frac{J}{\rho_l} \vec{n}  \right) \cdot \vec{n}.
$$

At the boundary ``liquid--cell'', the fluid flow velocity is zero, $\vec{v} = 0$,  due to the presence of viscous friction and no penetration through the substrate and cell wall. The following boundary conditions are used on a free surface of the liquid~\cite{Barash2009},
$$
	\frac{d \sigma}{d \tau} = \mu \left(\frac{\partial v_n}{\partial \tau} +    \frac{\partial v_{\tau}}{\partial n} -v_{\tau} \frac{\partial \phi}{\partial \tau} \right), \qquad p - P_{\nu} = P_L + 2 \mu \frac{\partial v_n}{\partial n},
$$
where $\vec \tau$ is a vector tangent to the surface, directed perpendicularly to the right when looking along the normal vector direction (the arrow above the directional derivative is omitted for brevity), $P_L = \sigma \left(1/R_1  + 1/R_2 \right)$ is the Laplace pressure, $R_1, \, R_2$ are the main radii of curvature on the free surface, $d \sigma/ d \tau =  \sigma^\prime \, \partial T/ \partial \tau$, $\phi$ is the angle between the vector $\vec \tau$ and $r$-axis (or angle between the vector $\vec n$ and $z$-axis)~\cite{Turchaninova2021}. Calculation details of $\partial \phi / \partial \tau$ are described in Ref.~\cite{Barash2009a}. In addition, on the axis of symmetry $v_r = 0$ and $\partial v_z / \partial r =0$~\cite{Turchaninova2021}.

The particle velocity on the free surface of the liquid and along the axis of symmetry is calculated as  $\vec{u} = \{ v_r, \, v_z - u_\mathrm{sed}   \}$, where $u_\mathrm{sed}= 2 r_p^2 (\rho_p - \rho_l)g/(9 \mu)$ is the particle sedimentation velocity.

At the ``liquid--cell'' boundary, the condition for particle reflection with conservation of kinetic energy takes the form of
$\vec{u} = \vec{u}_c - 2\vec{n} \cdot\left(\vec{n} \cdot \vec{u}_c   \right)$, where $\vec{u}_c$ is the particle velocity in collision with the boundary.

As the contact line is well away from the centre where the cluster is assembled in the cell, we use a simple boundary condition, $\theta = \pi / 2$, for the contact angle.

\subsection{Closing relations}

The vapor flux density is determined using the Hertz–-Knudsen formula, which takes into account the dependence of saturated vapor pressure on temperature~\cite{AlMuzaiqer2021},
\begin{equation}\label{eq:J}
	J = \alpha_\mathrm{evap} \sqrt{\frac{M}{2 \pi R T_\mathrm{sat}}} \left(P_\mathrm{sat}(T) - P_{\nu}  \right),
\end{equation}
where the coefficient of accommodation $\alpha_\mathrm{evap} = 9 \times 10^{-4}$ (the value of this parameter was chosen so that the evaporation time was close to the experimental value), $R$ is the universal gas constant, $P_\mathrm{sat}$ is the saturated vapor pressure. The saturated vapor pressure has been calculated using the empirical formula~\cite{AlMuzaiqer2021}
$$P_\mathrm{sat} = A_2 + \frac{A_1 - A_2}{1 +\exp\left((T_l - x_0)/\Delta x   \right)},$$ where $A_1 = - 1762.7$, $A_2 = 503914.8$, $x_0 = 383.2$, $\Delta x = 20.5$.

In this model the dependence of surface tension on temperature is described by the linear approximation $\sigma = \sigma_0 + \sigma^\prime (T_l - T_0)$.

\subsection{Numerical simulation}

\textit{COMSOL Multiphysics} (version 6.2) software was used to solve this problem. We have chosen the following modules: \textit{``Laminar flow''} for calculate the velocity and pressure fields of a single-phase liquid flow in the laminar flow mode, \textit{``Heat transfer in solids and fluids''} for simulation of heat transfer in solids and liquids by means of thermal conductivity, convection and radiation, and \textit{``Particle tracing for fluid flow''} for modeling and tracking the movement of micro- and macroscopic particles in the liquid.

The two-dimensional geometry of the axisymmetric surface was used, and several subdomains were built (the copper rod, the substrate and the liquid layer according to the Table~\ref{tab:GeometricParameters}). For the domain corresponding to the liquid layer, a moving mesh with a minimal grid cell size is defined (extremely fine). For the rest of the domains, the small value of the cell size is selected (fine). In addition, adaptive mesh refinement was used.

For the geometry corresponding to the rod, the material ``copper'' was selected from the materials library \textit{COMSOL Multiphysics}. For other domains, the specific heat capacity, thermal conductivity and density, as well as the dynamic viscosity for isopropyl alcohol, were prescribed manually. The values of all used parameters correspond to  Table~\ref{tab:PhysicalParameters}.

Two solvers were created for the simulation: \textit{Study1} for the calculation of hydrodynamics and \textit{Study2} for particle tracing. In \textit{Study1}  nonlinear method with a constant damping factor (is equal 0.9) was used, maximum number of iterations is equal 10, maximum time step 0.1~s. In \textit{Study2} highly nonlinear method with initial damping factor is $10^{-4}$ and minimum damping factor is $10^{-8}$ was used. Number of iterations is equal 250, maximum time step 0.5~s.

To simulate the rolling of particles on the surface of the cell bottom, a subdomain $\Omega_\mathrm{bot}$ was introduced near the bottom where gravity was disabled,  $\Omega_\mathrm{bot} \in [r=R_h..R_\mathrm{in};\, z= 0..h_\mathrm{cri}]$, where $h_\mathrm{cri} \ll h_0$. The value $h_\mathrm{cri}=d_p$ is used for calculations. This crude approach does not allow us to take into account the force of rolling friction, i.e. it is a zero approximation.

As a result of numerical calculations, the temperature field of the cell containing the liquid, the velocity field of fluid flow, and the particle tracks for different time points were constructed. All calculations were performed up to the time when the film thickness in any place reached a value close to the critical thickness $h_\mathrm{cri}$, since thermocapillary rupture of the film with the formation of a local dry spot should be expected later in time.

\section{Results and discussion}
\begin{figure}[h!]
	\centering
	\includegraphics[width=0.6\textwidth]{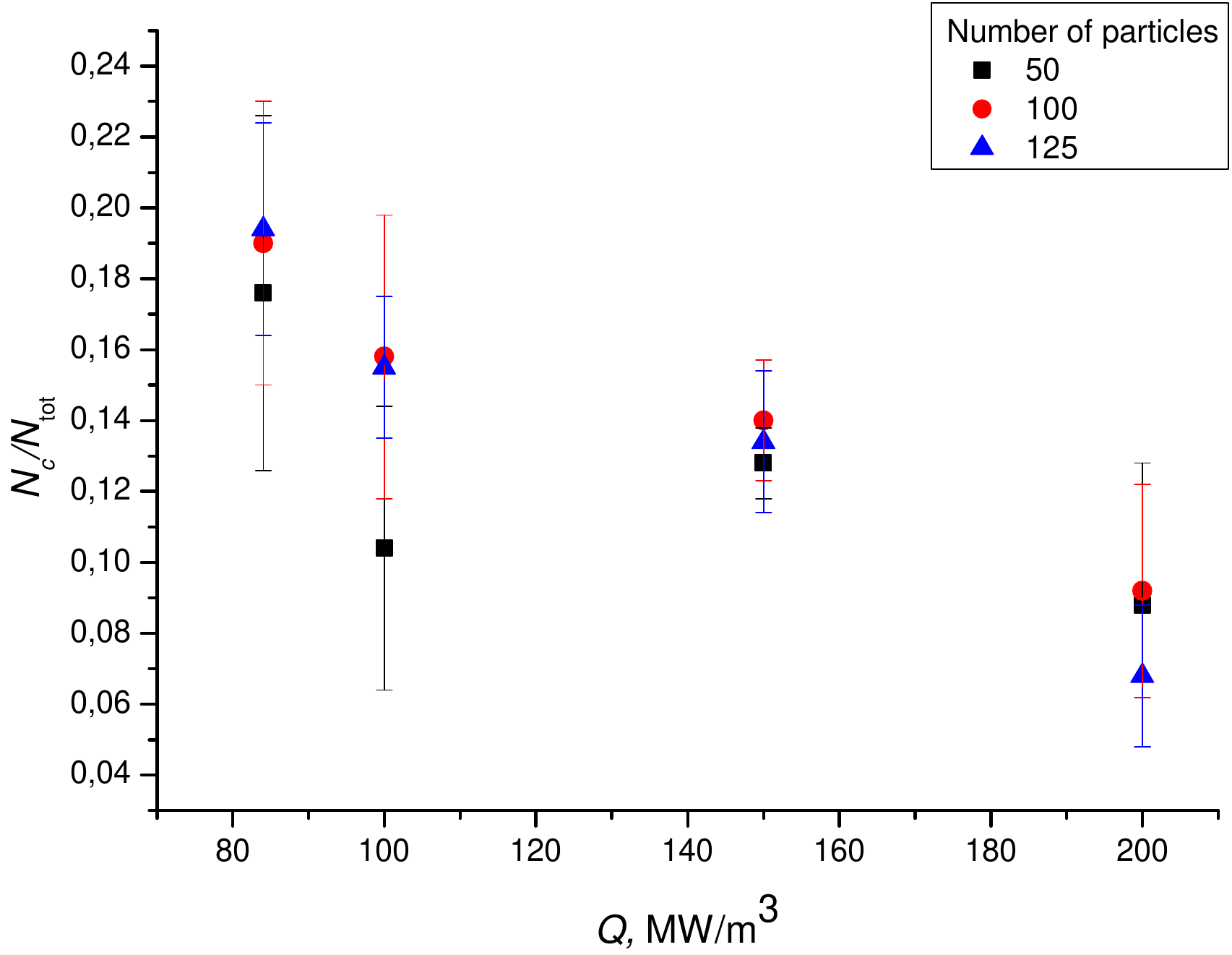}(a)\\
	\includegraphics[width=0.6\textwidth]{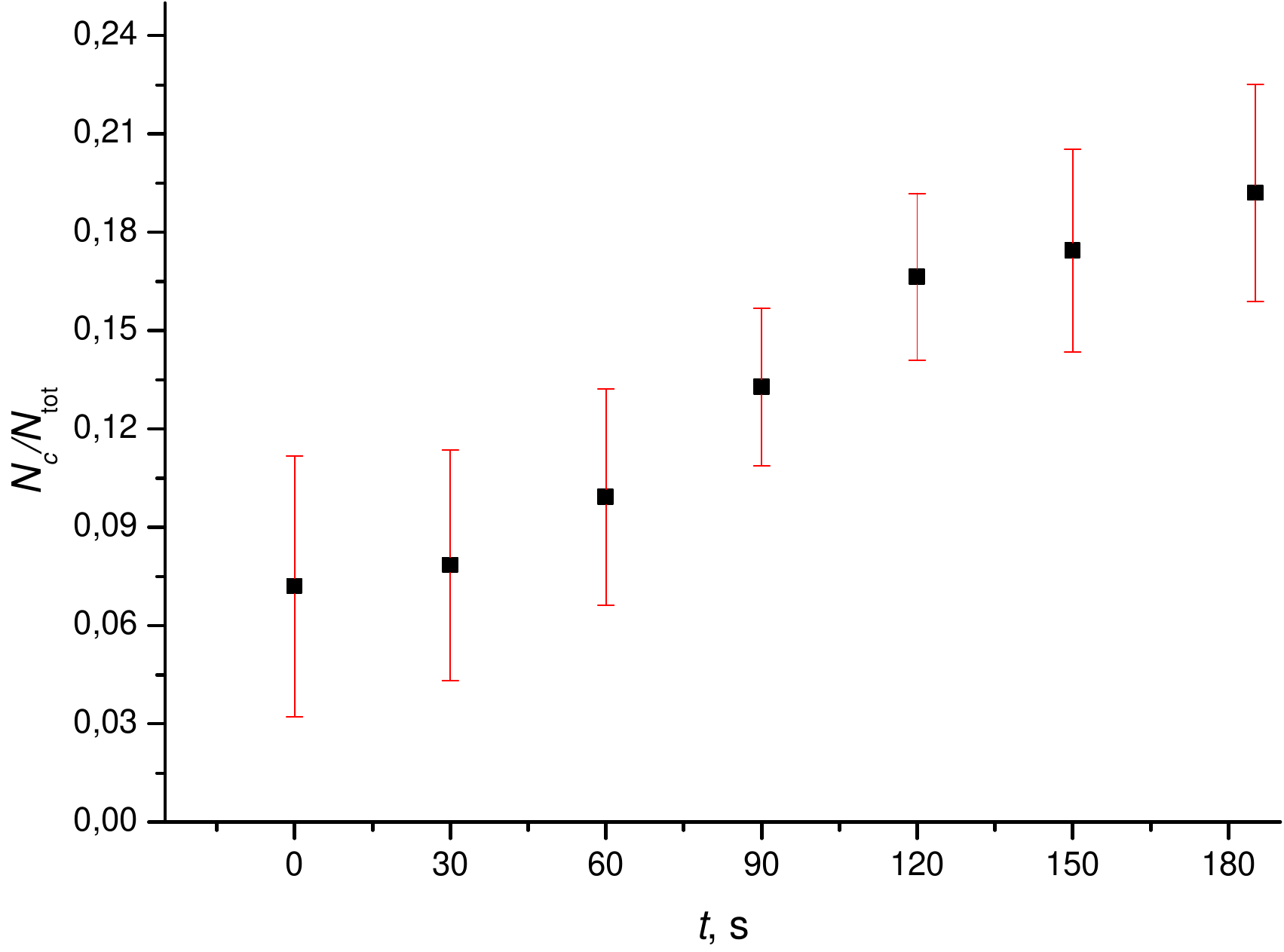}(b)
	\caption{Dependence of the particle fraction in the central cluster (a) on the volumetric heat flux and (b) on time ($Q=84$ MW/m$^3$, $N_\mathrm{tot}=125$).}
	\label{fig:NcNtotVsQ}
\end{figure}

A series of computational experiments have been carried out for different values of volumetric heat flux, $Q$, and the number of particles in the system, $N_\mathrm{tot}$. Each simulation of particle motion was repeated five times, after which the result was averaged over all trials. The standard deviation was used to calculate the error (the initial position of the particles was randomly generated), see Fig.~\ref{fig:NcNtotVsQ}. The temperature and flow velocity fields were calculated once for each considered value of the parameters $N_\mathrm{tot}$ and $Q$. The calculation time depends on $N_\mathrm{tot}$. It varies from a few hours to one or two days (Intel(R) Core(TM) i7-5820K, CPU 3.3~GHz, RAM 32 Gb, SSD 128~Gb (system), SSD 1~Tb (data)).

\begin{figure}[h!]
	\centering
	\includegraphics[width=0.6\textwidth]{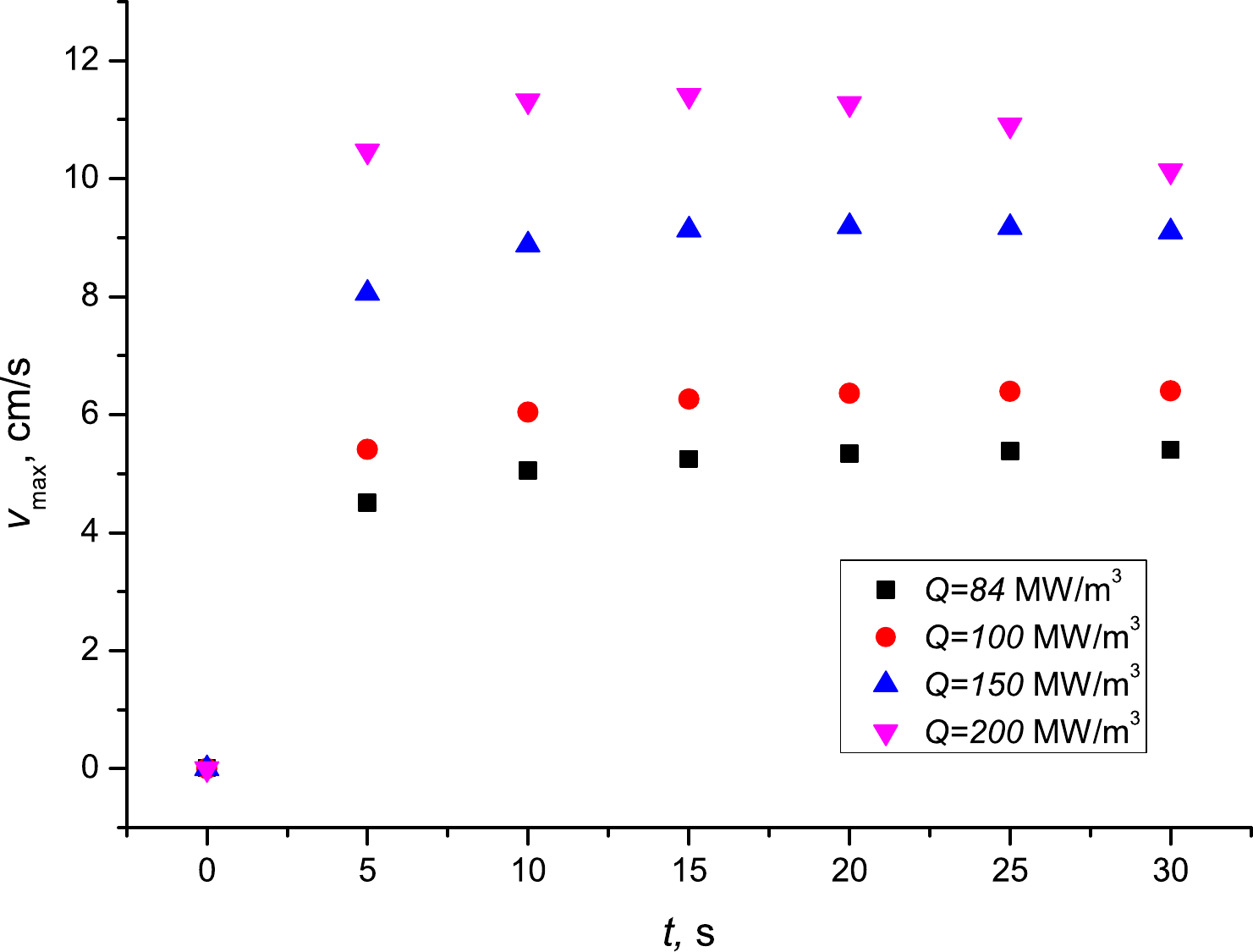}
	\caption{Dependence of the maximum thermocapillary flow velocity on time for several values $Q$ (the numerical calculation error is less than the marker size).}
	\label{fig:v_maxVsTimeInCellWithHeater}
\end{figure}

In the experiment~\cite{AlMuzaiqer2021}, a particle cluster was observed above a heated copper rod. Our calculations allow us to understand how the heating power affects the fraction of particles entering the cluster. In Fig.~\ref{fig:NcNtotVsQ}a, the dependence of $N_c / N_\mathrm{tot}$ on the value of the parameter $Q$ is shown. Here, $N_c$ is the particles number, which get into in the cluster. As $Q$ increases, the ratio $N_c / N_\mathrm{tot}$ decreases regardless of the value $N_\mathrm{tot}$ taken (50, 100 or 125). This is because the velocity of the thermocapillary flow depends on $Q$. For example, in Fig.~\ref{fig:v_maxVsTimeInCellWithHeater}, an increase in the value of maximum velocity, $v_\mathrm{max}$, is shown with an increase in volumetric heat flux. Only small concentrations are considered, so it is worth assuming that, perhaps, with a larger number of particles, the dynamics shown in Fig.~\ref{fig:NcNtotVsQ}a, will change. In order to make predictions for a case with a high concentration of particles, it is necessary to take into account the dependence of viscosity on concentration when constructing a model, since this affects hydrodynamics. Additionally, with a large number of particles in a cluster, a complex structure forms with many micropores, which are voids between packed microparticles. Filtration flows need to be taken into account within these pores. Liquid flows through these pores, while particles carried by the flow become stuck in this structure when they collide with it. Unfortunately, \textit{COMSOL Multiphysics} does not allow us to take into account particle size, as it is assumed that particles are points. Thus, our model does not allow us to predict the shape of the cluster. Perhaps the volume of particles can be mimicked by using some kind of interaction potential. Here, for simplicity, it is assumed that the particles don't interact. Another way to account for the volume of particles is to develop a calculation algorithm and program code, for example, as was done earlier when describing the evaporative self-assembly of particles in a drop drying on a substrate~\cite{Kolegov2019}.

\begin{figure}[h!]
	\centering
	\includegraphics[width=0.6\textwidth]{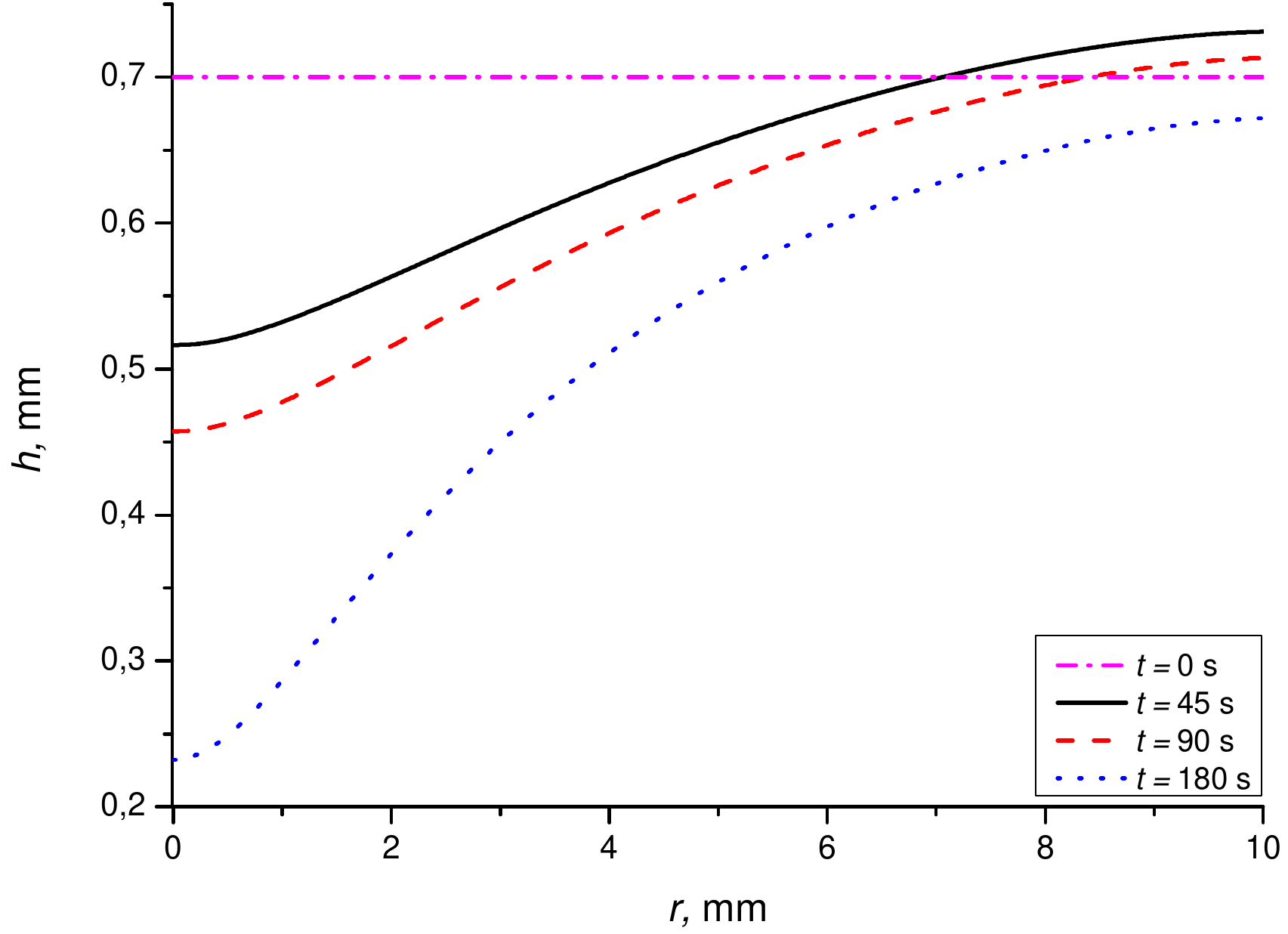}
	\caption{The height of the liquid layer in the cell at different points in time ($Q = $ 84 MW/m$^3$).}
	\label{fig:filmThicknessInCellWithHeater}
\end{figure}

The fraction of particles trapped in the central cluster increases over time (Fig.~\ref{fig:NcNtotVsQ}b). Whether a particle enters a cluster or continues its journey, carried away by a circulating Marangoni flow, depends on the balance of gravity, $F_g$, and drag force, $F_d$ (if we do not consider the buoyancy force, since $F_g > F_A$). In our opinion, this is one of the main mechanisms that influences the formation of the cluster and initiates this process.  However, in reality, other factors may also affect the further growth of the cluster. For example, changes in viscosity due to local increases in concentration, particle collisions, and friction between particles and the bottom of the cell during motion (rolling friction) can all contribute to the growth of the clusters.

\begin{figure}[h!]
	\centering
	\includegraphics[width=0.6\textwidth]{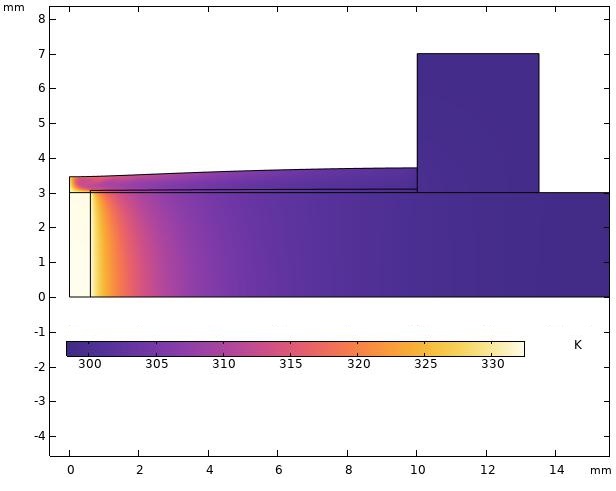} (a)\\
    \includegraphics[width=0.6\textwidth]{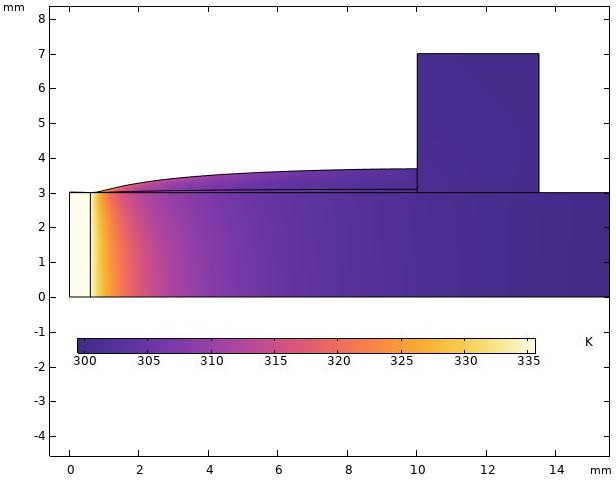} (b)
	\caption{Liquid and cell temperature field for (a)
$t =$ 90~s and (b) 185.4~s ($Q = $ 84 MW/m$^3$).}
	\label{fig:TemperatureInCellWithHeater}
\end{figure}

Let us estimate the critical value of the vertical component of flow velocity, which allows us to determine whether sedimentation or particle motion by the ascending flow prevails. A particle that falls into the region of the copper rod becomes part of the cluster if $F_g|_z > F_d|_z$. This condition can be written as $m_p g > (v_z - u_z) m_p / \tau_p$. When particles are transferred towards the heater along the bottom of the cell for a long time by fluid flow, their vertical velocity $u_z = 0$, so we get the inequality $\tau_p g > v_z$. A particle enters the cluster if the vertical component of the flow velocity $v_z$ is less than the velocity $\tau_p g$. In another case, the particle is caught in the ascending flow, and its circulation continues. Therefore, the higher the velocity $v_\mathrm{max}$ ($v_z \sim v_\mathrm{max}$) (Fig.~\ref{fig:v_maxVsTimeInCellWithHeater}), the greater the probability that a particle will escape from the cluster (Fig.~\ref{fig:NcNtotVsQ}a).

\begin{figure}
	\centering
	\includegraphics[width=0.6\textwidth]{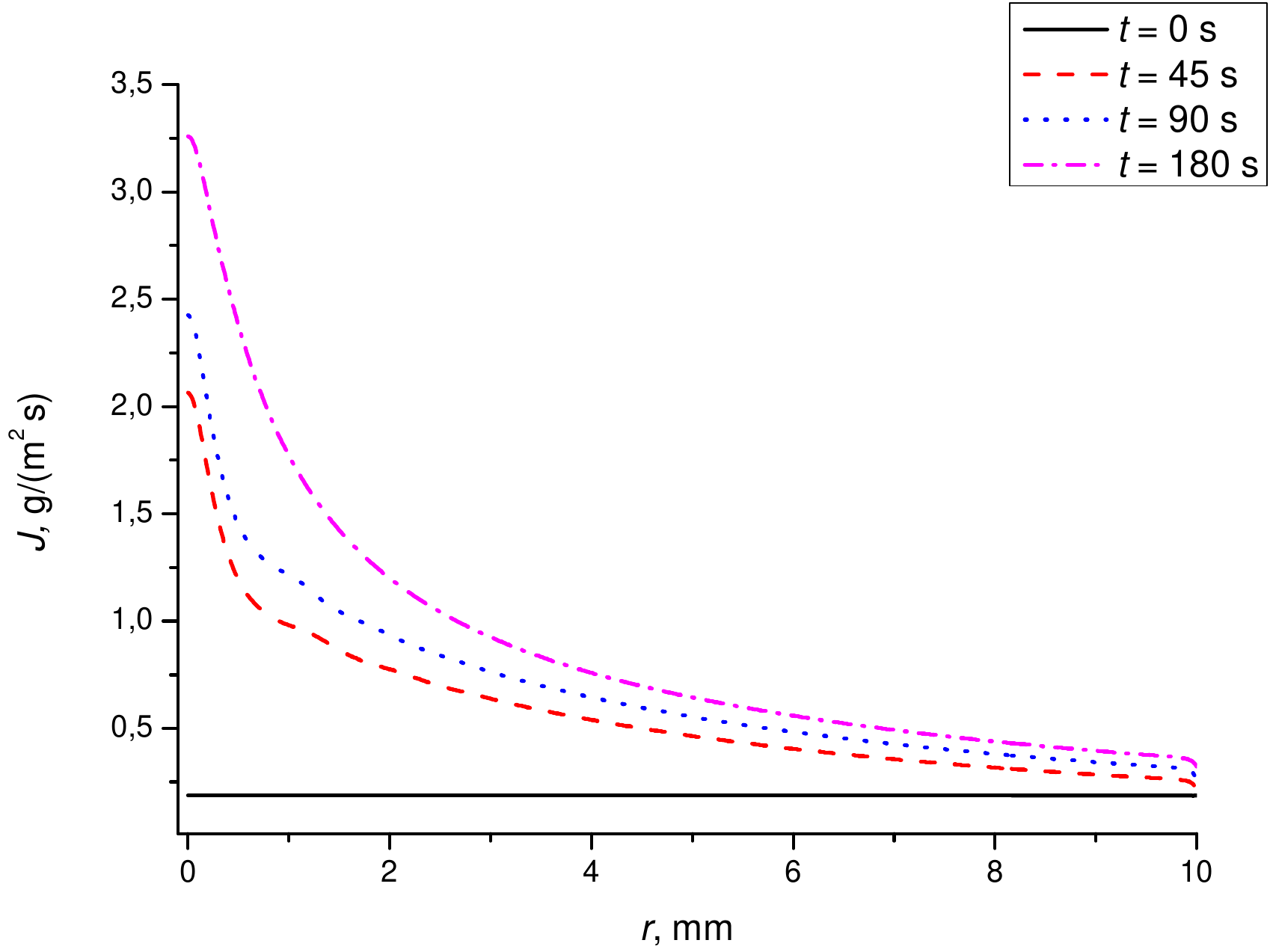}
	\caption{Local vapor flux density at several consecutive points in time ($Q = $ 84 MW/m$^3$).}
	\label{fig:evaporationFluxInCellWithHeater}
\end{figure}

The thermocapillary effect causes the free surface of the liquid to curve due to uneven heating (Fig.~\ref{fig:filmThicknessInCellWithHeater}). It is in qualitative agreement with the results from the 1D model~\cite{AlMuzaiqer2021}. This is due to a surface tension gradient caused by a temperature difference. By the cell wall, the liquid layer becomes higher as the liquid flows there, because the surface tension is higher there due to a lower temperature (Fig.~\ref{fig:TemperatureInCellWithHeater}). In the central area of the cell, the thickness of the liquid layer decreases, which leads to thermocapillary rupture and the formation of a dry spot (Fig.~\ref{fig:TemperatureInCellWithHeater}). In the area of the heater, there is an additional decrease in the thickness of the liquid layer due to uneven evaporation. This is because the higher the temperature of the liquid, the more actively it evaporates (Fig.~\ref{fig:evaporationFluxInCellWithHeater}). Over time, as the liquid evaporates, its height decreases (Fig.~\ref{fig:filmThicknessInCellWithHeater}). At $Q = $ 84 MW/m$^3$, the temperature difference $\Delta T$ is approximately 34~K. Let us determine the value of the Marangoni number, $\mathrm{Ma}=-\sigma^\prime \Delta T\, h_0/(\mu D_T)\approx 1.258\times 10^{4}$, where the thermal diffusivity coefficient is $D_T = \kappa_l / (\rho_l c_l)\approx 6.3\times 10^{-8}$ m$^2$/s. At such relatively large temperature differences considered here, strong thermocapillary convection arises, which is described by the proposed model. However, it should be noted that the temperature must not exceed a critical value at which liquid boiling will occur. Boiling can have a negative impact on particle self-assembly. Furthermore, the model does not describe such a process.

\begin{figure}[h!]
	\centering
	\includegraphics[width=0.6\textwidth]{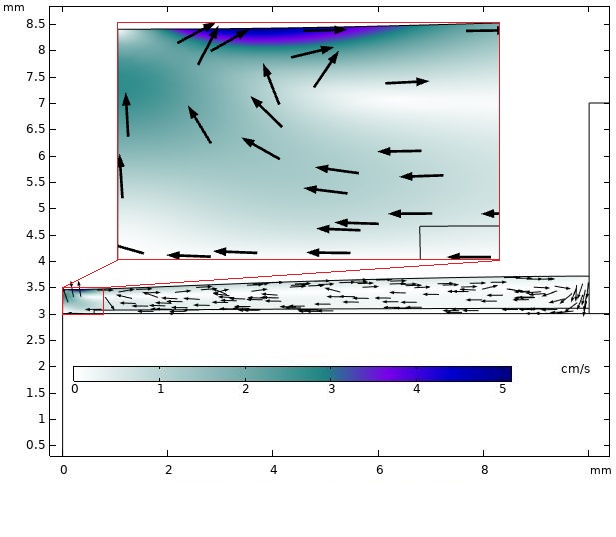} (a)\\
	\includegraphics[width=0.6\textwidth]{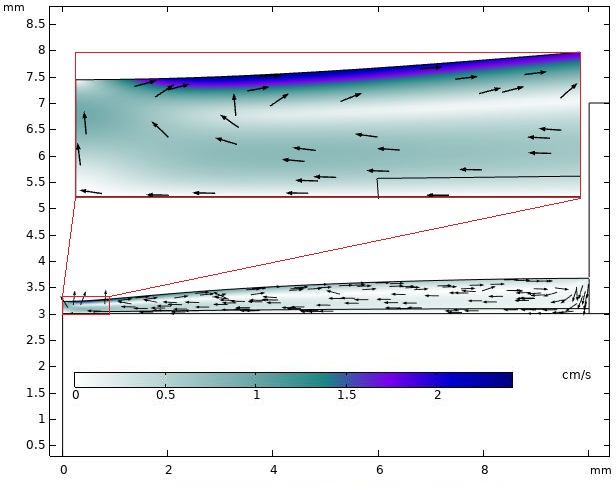} (b)
	\caption{Velocity field of the fluid flow for (a)
		$t =$ 90~s and (b) 180~s ($Q = $ 84 MW/m$^3$).}
	\label{fig:VelocityFieldInCellWithHeater}
\end{figure}

\begin{figure}[h!]
	\centering
	\includegraphics[width=0.99\textwidth]{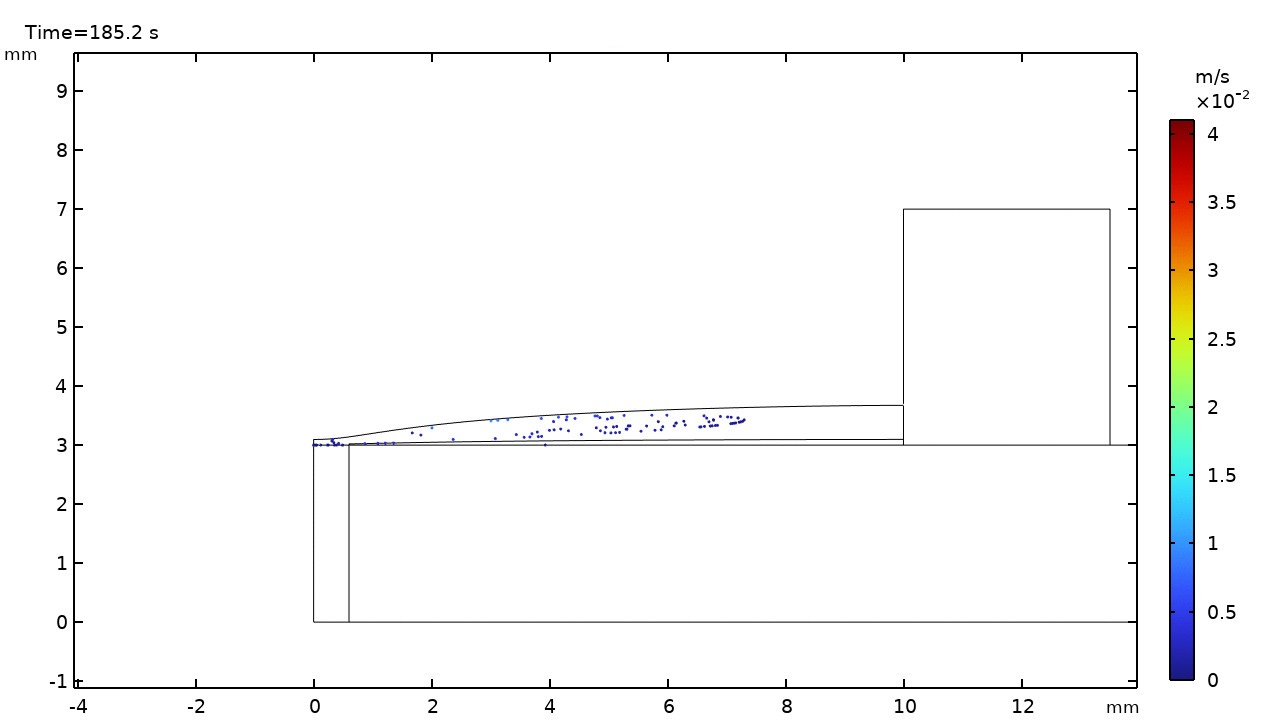}
	\caption{Particle dynamics at $Q = $ 84 MW/m$^3$ and $N=125$ (multimedia available in Supplementary Material).}
	\label{fig:ParticleMotion}
\end{figure}

The velocity field of a fluid flow is shown in Fig.~\ref{fig:VelocityFieldInCellWithHeater}. The thermocapillary flow is directed along the free surface from the center to the cell wall, since the surface tension gradient is positive (Fig.~\ref{fig:VelocityFieldInCellWithHeater}). Along the bottom of the cell, the flow is directed from the wall towards the heater. This circulating flow leads to the transfer of microspheres (Fig.~\ref{fig:ParticleMotion}, multimedia available online). The peak flow velocity is attained at the free surface of the liquid layer within the region adjacent to the heater, where the temperature gradient at the ``liquid--air'' boundary reaches its maximum.

To understand which of the two processes, evaporation or thermocapillary flow, has a greater influence on the thinning of the liquid layer near the heater, let us compare two characteristic times. We estimate the characteristic evaporation time as $t_e = h_0 / v_e$, where the characteristic evaporation velocity is $v_e = J_\mathrm{max} / \rho_l$ (for $Q=$ 84~MW/m$^3$, $J_\mathrm{max}\approx 3.25 \times 10^{-3}$ kg/(m$^2$s) according to Fig.~\ref{fig:evaporationFluxInCellWithHeater}). The characteristic time of thermocapillary convection can be estimated by the formula $t_M = h_0 / v_M$, where the characteristic Marangoni flow velocity is $v_M \approx 2.5\times 10^{-2}$ m/s (Fig.~\ref{fig:VelocityFieldInCellWithHeater}b). Substituting the parameter values, we obtain $t_e \approx 169$ s and $t_M \approx 0.03$ s. The rapid thinning of the liquid film in the central part of the cell is associated with thermocapillary convection, since $t_M \ll t_e$. A dry spot forms as a result of a thermocapillary rupture, where the central cluster ends up outside the liquid. Subsequent evaporation of the remaining liquid from the cell and the transport of particles suspended in it do not affect the size of this cluster. Let us also estimate the thermal relaxation time, $t_r = h_0^2 / D_T \approx$ 7.8 s. From the relation $t_M \ll t_r$, it follows that heat transfer in the liquid due to convection dominates over thermal conduction.

\section{Conclusion}
The control of thermocapillary assembly of clusters of colloidal particles is important for applications as diverse as the creation of photonic crystals for micro- and optoelectronics, the formation of membranes for biotechnology, and surface cleaning for lab-on-a-chip devices. A comprehensive understanding of the primary mechanisms governing the formation of such clusters, as observed in experimental studies~\cite{AlMuzaiqer2021126550,AlMuzaiqer2021}, is essential.

This article considers a two-dimensional mathematical model describing the transfer of particles by a thermocapillary flow in an unevenly heated cell during the evaporation of the liquid. The previously proposed one-dimensional model only took into account the transport of particles near the substrate~\cite{AlMuzaiqer2021}. But in reality, some particles can also circulate in the liquid volume due to the upward flow in the area of the heater. In Ref.~\cite{AlMuzaiqer2021}, particle transport was described using the convection-diffusion equation. The motion of each individual particle is explicitly modeled here. This allowed us to study one of the main mechanisms that initiates the formation of particle clusters. Whether a particle circulates with the flow or stops in the area of the heater, initiating the formation of a cluster, depends on the ratio of gravity and drag force. The results of numerical calculations showed that, with an increase in the volumetric heat flux density, $Q$, the fraction of particles entering the cluster decreased. This phenomenon occurs because an increase in $Q$ enhances the thermocapillary flow, thereby reducing the likelihood of particles entering the cluster.

It is important to note that this model has been constructed to work with a small number of particles. The viscosity of the liquid may vary depending on local concentration. In addition, in order to build more accurate models in the future, it is necessary to take into account the interaction of particles with each other, for example, their collision, and the interaction ``particle-substrate'' such as rolling friction. To describe the shape of a cluster, the volume of particles must be taken into account (two particles cannot occupy the same space at the same time, even partially). It is probably worth paying attention to the fact that as a result of the accumulation of particles, a system of holes is formed through which the liquid is filtered.

\section*{Data Availability}
The data that support the findings of this study are available from the corresponding author upon reasonable request.

\section*{Conflicts of Interest}
The authors declare that they have no known competing financial interests or personal relationships that could have appeared to influence the work reported in this paper.

\begin{acknowledgments}
This work is supported by Grant No. 22-79-10216 from the Russian Science Foundation (\href{https://rscf.ru/en/project/22-79-10216/}{https://rscf.ru/en/project/22-79-10216/}).
\end{acknowledgments}

\bibliography{Kondrashova2024}

\end{document}